\documentclass[prd, preprint, 11pt]{revtex4-1}

\usepackage{amsmath}
\usepackage{amssymb}
\usepackage{setspace}
\usepackage{graphicx}
\usepackage{natbib}
\usepackage{float}

\begin{document}
 
 %

\begin{center}
 { \large {\bf How the quantum emerges from gravity}}


\vskip 0.2 in

{\large{\bf Anushrut Sharma$^{*}$ and Tejinder P.  Singh$^{\dagger}$}}

{\it $^{*}$Indian Institute of Technology Bombay, Powai, Mumbai 400076, India}\\  
{\it $^{\dagger}$Tata Institute of Fundamental Research,}
{\it Homi Bhabha Road, Mumbai 400005, India}\\
{\tt email addresses: anushrut@iitb.ac.in, tpsingh@tifr.res.in}\\

\end{center}

\bigskip

\centerline{\bf ABSTRACT}
\noindent The dynamics of a massive, relativistic spinning particle could be described either by the Dirac equation or by the Kerr solution of Einstein equations. However, one does not know a priori as to which of the two systems of equations should be used in a given situation, and the choice is dictated by experiments. It is expected that the Dirac equation holds for microscopic masses, and the Kerr solution for macroscopic masses. This suggests that Einstein gravity and the Dirac theory are limiting cases of a common underlying theoretical framework. Here we propose  that such a framework is provided by a geometric theory of gravity on a Riemann-Cartan spacetime, which includes torsion. The Dirac equation emerges as the torsion dominated, gravity-free limit of this framework.

\vskip 1 in

\centerline{March 31, 2014}

\bigskip

\centerline{\it This essay received an honorable mention in the Gravity Research Foundation 2014 Essay Contest}

\bigskip

\bigskip


\setstretch{1.1}
\newpage

\noindent The dynamics of a relativistic spinning elementary `particle` of mass $m$ is described by the Dirac equation. Or (keeping in view also the spacetime geometry) it could be described by the Kerr solution of Einstein equations. Which of the two? We do not know a priori, except through results of experiments and astronomical observations. We  naively expect the answer to be Dirac equation if $m\ll m_{Pl}$ and Kerr geometry if $m\gg m_{Pl}$, where $m_{Pl}$ is Planck mass $\sim 10^{-5}$ grams. However, the scale $m_{Pl}$ appears neither in Einstein equations nor in the Dirac equation, strongly suggesting that there should be an underlying theory which does involve $m_{Pl}$, and to which the Dirac equation/Einstein equations are the small mass/large mass approximation, respectively. The purpose of the present essay is to propose the idea for such an underlying framework, in the form of a (gravity + torsion) theory on a Riemann-Cartan spacetime. Quantum theory in the form of Dirac equation emerges from the underlying theory when torsion dominates gravity. The Dirac quantum state is identified with a complex torsion \cite{Sharma2014}.

A significant motivation for this specific form of the underlying theory comes from noting the following contrast amongst the symmetry groups that underlie the Dirac equation and Einstein equations. Elementary quantum particles are represented by irreducible unitary representations of the Poincare group (which includes both Lorentz boosts and translations), and are labeled by mass and spin. On the other hand, in general relativity, the structure group acting on the tangent space of a manifold is the Lorentz group, and translations are not included. This intriguing disparity is resolved by introducing torsion, which restores the Poincare group in relativistic gravity, because torsion bears a relation to translations similar to the relation between curvature and linear homogeneous transformations \cite{Trautman}. Bringing quantum theory and gravity into a common framework seems more tangible then, if the gravity theory includes torsion. However, torsion tends to get ignored in theoretical investigations, and naturally so, because there is little observational evidence for it. Our essay sheds new insight on how torsion can be retained, and used to provide a bridge from the Dirac equation to Einstein gravity.  

Einstein gravity, treated as a metric theory of gravity, and expressed as
\begin{equation}
G_{\mu\nu} +\Lambda g_{\mu\nu} = \frac{8\pi G}{c^4} T_{\mu\nu}
\end{equation}
looks very different from the Dirac equation
\begin{equation}
i\hbar\gamma^{\mu}\partial_{\mu}\psi = mc\psi
\end{equation} 

However the two theories start to look strikingly similar in the Newman-Penrose tetrad formalism
\cite{Newman1962, Newman1963}, in which the fundamental quantities are not the metric, but a tetrad of four null vectors, $\bf l$, $\bf n$, $\bf m$ and $\bf\bar{m}$, where the former two are real, and the latter two are complex conjugates of each other. The directional derivatives associated with the tetrads are denoted by the special symbols (equality stands for corresponding derivative)
\begin{equation}
D={\bf l}, \quad \Delta={\bf n}, \quad \delta = {\bf m}, \quad \delta^{*} = \overline{\bf m}
\end{equation}
Via the definitions of the covariant derivatives of the tetrads, there arise in the formalism twelve important complex quantities, known as spin-coefficients, and denoted by the symbols \cite{Chandra}
\begin{equation}
 \kappa, \sigma, \lambda, \nu, \rho, \mu, \tau, \pi, \epsilon, \gamma, \alpha, \beta
 \end{equation}  
 The twenty independent components of the Riemann tensor include the ten Ricci tensor components, described by the four real and three complex quantities
  \begin{equation}
 \Phi_{00}, \Phi_{11}, \Phi_{22}, \Lambda, \Phi_{02}, \Phi_{20}, \Phi_{01}, \Phi_{10}, \Phi_{12}, \Phi_{21}
 \label{Ricci}
 \end{equation}
 and the ten Weyl tensor components described by the five complex quantities
 \begin{equation}
 \Psi_0, \Psi_1, \Psi_2, \Psi_3, \Psi_4
 \end{equation}
 
 The Riemann tensor can now be elegantly expressed in terms of the spin-coefficients and their derivatives, via eighteen complex first order equations, known as the Ricci identities. These equations typically take the form \cite{Chandra}
 \begin{equation}
D\rho - \delta^{*}\kappa = (\rho^2+\sigma\sigma^{*}) + \rho( \epsilon + \epsilon^{*})
 -\kappa^{*}\tau -\kappa (3\alpha +\beta^{*}-\pi) + \Phi_{00}
 \end{equation}
 with every equation involving a pair of derivatives of the spin-coefficients, products of spin-coefficient pairs, and components of the Ricci / Weyl tensor. There being thirty-six such real equations, and there  being only twenty independent components of Riemann, the eighteen Ricci identities obey sixteen constraints, known as eliminant conditions. The Ricci components in these equations are provided by the chosen field theory of gravity, for instance, Einstein gravity.
 
 Let us now compare the Ricci identities with the Dirac equations in a curved spacetime, written as four equations for the pairs of spinors $F_1$, $F_2$, $G_1$ and $G_2$, in the Newman-Penrose formalism \cite{Chandra}
\begin{equation}
(D+\epsilon - \rho) F_1 + (\delta^{*} + \pi - \alpha) F_2 = i\mu_{*} G_{1}
\label{D1}
\end{equation}
\begin{equation}
(\Delta + \mu - \gamma) F_2 + (\delta + \beta - \tau) F_1 = i\mu_{*} G_{2}
\label{D2}
\end{equation}
\begin{equation}
(D + \epsilon^{*} - \rho^{*}) G_{2} - (\delta + \pi^{*} - \alpha^{*}) G_{1} =i \mu_{*} F_{2}
\label{D3}
\end{equation}
\begin{equation}
(\Delta + \mu^{*} - \gamma^{*}) G_{1} - (\delta^{*} + \beta^{*} - \tau^{*}) G_{2} = i \mu_{*} F_{1}
\label{D4}
\end{equation}
where $\mu_{*} = mc/\sqrt{2}\hbar$.
If the four spinor components were to be suitably identified with four of the spin-coefficients, and if the mass terms on the right hand side of the Dirac equations are taken to be proportional to suitable components of the Ricci and Weyl, the formal structure of the Dirac equations becomes identical to the Ricci identities. Thus we propose the identification \cite{Sharma2014}
\begin{equation}
F_1 =\frac{1}{\sqrt{l_p}}\; \lambda, \quad F_2=-\frac{1}{\sqrt{l_p}}\;\sigma, \quad G_1=\frac{1}{\sqrt{l_p}}\;\kappa^{*}, \quad G_2=\frac{1}{\sqrt{l_p}}\;\nu^{*}
\label{match}
\end{equation}
and since we are interested in recovering the flat spacetime limit of the Dirac equations from the Ricci identities, we set the remaining eight spin cefficients to zero:
\begin{equation}
\rho=\mu=\tau=\pi=\epsilon=\gamma=\alpha=\beta=0
\label{vanishspin}
\end{equation}
These are precisely the eight coefficients which appear explicitly in the Dirac equations 
(\ref{D1}) - (\ref{D4}). It is remarkable that there is a complementarity with general relativity [GR]: all black hole solutions in GR belong to Petrov type D and can be constructed in a tetrad frame in which these eight spin-coefficients are non-zero, whereas those of (\ref{match}) vanish \cite{Chandra}.

By taking appropriate combinations of pairs of Ricci identities, using the correspondence (\ref{match}), and identifying the Riemann components with the right hand sides of the four Dirac equations one can recover the Dirac equations  from the Ricci identities. However, the sixteen eliminant conditions which constrain the Ricci identities must be accounted for, and this results in undesirable constraints on the Dirac spinor components, thereby rendering this first attempt unsuccessful \cite{Sharma2014}.

However, there is a physical situation when the Riemann tensor has exactly thirty-six independent components, and there are no eliminant conditions to be imposed on the Ricci identities. This is when spacetime has torsion [Riemann-Cartan spacetime], torsion being the antisymmetric part of the connection. This ties well with there being eighteen complex Ricci identities, and now the recovery of Dirac equations from the Ricci identities proceeds smoothly.  

The connection is now given by the general form
\begin{equation}
\Gamma_{\mu\nu}^{\ \ \lambda} = \left\{ _{\mu\nu}^{\ \ \lambda} \right\} - K_{\mu\nu}^{\ \ \lambda}
\end{equation}
where $\left\{ _{\mu\nu}^{\ \ \lambda} \right\}$ is the Christoffel symbol of the second kind, and  $K_{\mu\nu}^{\ \ \lambda}$ is the contortion tensor signifying the presence of torsion.
The Ricci tensor is not symmetric, and has six additional components, described by  the three complex quantities ($\Phi_0, \Phi_1$ and $\Phi_2$).  The Weyl tensor has ten additional components, described by the real quantities ($\Theta_{00}, \Theta_{11}, \Theta_{22}, \chi$) and the complex quantities ($\Theta_{01}, \Theta_{02}, \Theta_{12}$) \cite{Jogia}.

Moreover, the spin coefficients now have an additional term due to torsion, which we denote as
\begin{equation}
\kappa = \kappa^{\circ} +\kappa_1, \qquad \rho = \rho^{\circ} + \rho_1, \ \ \ \ \ \ \text{etc}.
\end{equation}
where the first term corresponds to the torsion free part and the second term corresponds to the torsion component of the spin coefficients. 
The Ricci identities are also modified when torsion is included, and a typical example is \cite{Jogia}
\begin{multline}
D\rho - \delta^{*}\kappa = \ \rho (\rho + \epsilon + \epsilon^{*}) + \sigma \sigma^{*} - \tau \kappa^{*} - \kappa (3\alpha + \beta^{*} - \pi) + \Phi_{00}   \\
- \rho(\rho_1 - \epsilon_1 + \epsilon_{1}^{*} ) - \sigma \sigma_1^{*} + \tau \kappa_1^{*} + \kappa(\alpha_1 + \beta_{1}^{*} - \pi_1 ) + i\Theta_{00}
\end{multline}

Working under the conditions (\ref{vanishspin}) the Ricci identities can be solved for the components of the Riemann in terms of the four non-vanishing spin-coefficients and their derivatives, and in particular they can be solved when the torsion-free (symmetric) part is exactly zero. We call this the gravity-free, torsion dominated limit. This is the Minkowski flat Rieman-Cartan spacetime with torsion and the one of interest to us here. Now the connection equals the contortion tensor, i.e.  $\kappa=\kappa_1$, $\sigma=\sigma_1$, $\lambda=\lambda_1$ and $\nu=\nu_{1}$ are the only 
non-zero spin-coefficients. Assuming the correspondence (\ref{match}) and using appropriate pairwise combinations of Ricci identities, the Dirac equations follow from the identities, provided the Riemann tensor satisfies the following four conditions \cite{Sharma2014} : 
\begin{equation}
\Phi_{20} + i\Theta_{20} + \Phi_{01} + i\Theta_{01} - \Psi_1 - \Phi_0  = i\mu_*\kappa^* 
\label{dir1}
\end{equation}
\begin{equation}
\Phi_{21} + i\Theta_{21} + \Phi_2 - \Psi_3 + \Phi_{02} + i\Theta_{02}  = i\mu_* \nu^* 
\label{dir2}
\end{equation}
\begin{equation}
i\Theta_{12} - \Phi_{12} + i\Theta_{00} - \Phi_{00} + \Phi_{2}^* -\Psi_{3}^* = i\mu_* \sigma 
\label{dir3}
\end{equation}
\begin{equation}
i\Theta_{10} - \Phi_{10} - \Phi_{0}^* - \Psi_{1}^* + i\Theta_{22} - \Phi_{22} = i\mu_* \lambda 
\label{dir4}
\end{equation}
By virtue of the solutions of the Ricci identities, these in fact are the same as the Dirac equations 
(\ref{D1}) - (\ref{D4}). What they amount to is a new set of field equations, different from Einstein gravity, wherein the Riemann is determined by a mass term proportional to a complex torsion on a Minkowski flat spacetime, and the complex torsion is conceptually the same object as the Dirac quantum state.

What have we achieved thus? We are proposing that a Riemann-Cartan geometry which includes gravity as well as torsion is more fundamental than both Einstein gravity and Dirac theory. It restores the fundamental status of the Poincare group across the board. Through the Ricci identities, the Riemann tensor is expressed in terms of the spin-coefficients and their derivatives, and the Ricci and Weyl are to be determined by a choice of equations of motion. One extreme is the torsion-free, gravity dominated limit, namely Einstein gravity, which presumably is the $m\gg m_{Pl}$ case, which involves $G$ but no $\hbar$. The other extreme is the  gravity-free, torsion dominated limit, namely the Dirac limit [Eqns. (\ref{dir1}) - (\ref{dir4})], presumably the case $m\ll m_{Pl}$, which involves $\hbar$ but not $G$. Thus, once we accept to identify a complex torsion with the Dirac quantum state, the Ricci identities for a Riemann-Cartan spacetime become a common source for Einstein gravity and Dirac equations. The Dirac equation, and in this sense quantum theory itself, is seen as an emergent geometry where complex torsion lives on a Minkowski flat spacetime.

In the domain between Einstein gravity and Dirac theory, lies unchartered territory, where 
$m\sim m_{Pl}$, and both $G$ and $\hbar$ make their appearance. The source for the Riemann tensor is a combination of the matter energy-momentum tensor and the complex torsion term of the type $imc\kappa/\hbar$, with the latter being possibly related to intrinsic spin.  The theory bears resemblance to the Einstein-Cartan-Sciama-Kibble theory (for a review see \cite{Hehl}), with one important difference: $\hbar$ is now explicitly present in the field equations, and torsion is complex. It is possible that the non-relativistic limit of the theory is a non-linear Schrodinger equation which might help verify/rule out the hypothesis that the collapse of the wave-function during a quantum measurement is caused by gravity. This particular feature is amenable to currently ongoing experimental tests \cite{RMP:2012} and its experimental investigation serves also as a test for the idea proposed in this essay. 

During the last century, many eminent physicists have emphasized the highly restrictive nature of a symmetric connection, and the possible fundamental significance that the skew-symmetric part of the connection (torsion) might hold. Noteworthy amongst them is Schrodinger, who highlights this aspect essentially throughout his insightful book {\it Space-time structure} \cite{Schrodinger1960} and in particular suggests on p. 115-16 the possibility of a geometric, torsion-oriented description of matter (see also Sciama \cite{Sciama1961}). Where we have added a new aspect is in suggesting a complex torsion, cast in the modern language of the Newman-Penrose formalism, and this complex feature permits emergence of quantum theory in a manner not quite feasible for a real (fully classical) geometric theory.   In the light of current ongoing experiments on gravity induced wave-function collapse, this family of ideas is worth revisiting.

The work of TPS is supported by a grant from the John Templeton Foundation [\# 39530].

\newpage

\centerline{\bf REFERENCES}

\bibliography{biblioqmts3}

\end{document}